\documentclass[12pt]{article}
\usepackage{graphics} 
\def\lsim{\mathrel{\lower2.5pt\vbox{\lineskip=0pt\baselineskip=0pt
          \hbox{$<$}\hbox{$\sim$}}}}
\def\gsim{\mathrel{\lower2.5pt\vbox{\lineskip=0pt\baselineskip=0pt
          \hbox{$>$}\hbox{$\sim$}}}}

\begin{document}   
\markright{Traversable wormholes...}
\def\Barcelo{Barcel\'o}
\title{Traversable wormholes from massless conformally coupled scalar fields}
\author{Carlos \Barcelo\ and Matt Visser\\[2mm]
{\small \it 
Physics Department, Washington University, 
St.~Louis, Missouri 63130-4899, USA.}}
\date{{\small 9 August 1999; \LaTeX-ed \today}}
\maketitle
\begin{abstract}
The massless conformally coupled scalar field is characterized by the
so-called ``new improved stress-energy tensor'', which is capable of
classically violating the null energy condition. When coupled to
Einstein gravity we find a three-parameter class of exact
solutions. These exact solutions include the Schwarzschild geometry,
assorted naked singularities, and a large class of traversable
wormholes.

\vspace*{5mm}
\end{abstract}
\def\Box{\nabla^2}
\def\SIZE{1.00}
\section{Introduction}
\label{i}

The equations of general relativity relate the geometry of a spacetime
manifold to its matter content. Given a geometry one can find the
distribution of mass-energy and momenta that support it. Unless there
are restrictions on the features that the matter content can possess,
general relativity allows the existence of spacetime geometries in
which apparently distant regions of space are close to each other
through wormhole connections. The existence or not of traversable
wormhole geometries has many implications that could change our way of
looking at the structure of spacetime (see \cite{visser1} for a survey
and bibliography on the subject).

The analysis of M.~Morris and K.~Thorne \cite{morris1} regarding
traversable wormhole geometries showed that, in order for the flaring
out of geodesics characteristic of a Lorentzian wormhole throat to
happen, it is necessary that the matter that supports the wormhole
throat be peculiar: it has to violate the Null Energy Condition (NEC)
\cite{hawking}. That is to say, even a null geodesic observer 
would see negative energy densities on passing the throat.  This
analysis was originally done with static spherically symmetric
configurations, but the NEC violation is a generic property of an
arbitrary wormhole throat \cite{visser2}.

It is often (mistakenly) thought that classical matter always
satisfies NEC. By contrast, in the quantum regime there are well-known
situations in which NEC violations can easily be obtained
\cite{visser1}.  For this reason, most investigations regarding
wormhole physics are developed within the realm of semiclassical
gravity, where the expectation value of the quantum energy-momentum
tensor is used as the source for the gravitational field
\cite{hochberg2}. 

However, one can easily see that the energy-momentum tensor of a
scalar field conformally coupled to gravity can violate NEC even at
the classical level \cite{flanagan}.  We would like to highlight that
this energy-momentum tensor is the most natural one for a scalar field
at low energies (energies well below the Planck scale, or any other
scale below which a scalar field theory might become
non-renormalizable), because its matrix elements are finite in every
order of renormalized perturbation theory \cite{callan}. In flat
spacetime, this energy-momentum tensor defines the same four-momentum
and Lorentz generators as that associated with a minimally coupled
scalar field and, in fact, it was first constructed as an improvement
of the latter \cite{callan}. Thus, we wish to focus attention on the
so-called ``new improved energy-momentum tensor'' of particle physics.
At higher energies, still well below the Planck scale, there may also
be other forms of classical violations of the NEC, such as higher
derivative theories \cite{hochberg,ghoroku}, Brans--Dicke theory
\cite{brans-dicke,anchordoqui}, or even more exotic
possibilities.\footnote{
For instance, we mention the work by H.~Ellis
\cite{ellis} in which he considered changing the sign in front of the
energy-momentum tensor for a minimally coupled scalar field. Reversing
this sign from the usual one, he found classical wormhole solutions,
which with hindsight is not surprising since reversing the
energy-momentum tensor explicitly violates the energy conditions. This
paper is of particular interest since it pre-dates the Morris--Thorne
analysis by 15 years.}

In this paper, we will concentrate on the massless conformally coupled
scalar field. We will explicitly show that it can provide us with the
flaring out condition characteristic of traversable wormholes. We have
analytically solved the Einstein equations for static and spherically
symmetric geometries. We find a three-parameter class of exact
solutions. These solutions include the Schwarzschild geometry, certain
naked singularities, and a collection of traversable
wormholes. However, in all these wormhole geometries the effective
Newton's constant has a different sign in the two asymptotic
regions. At the end of the paper we will briefly discuss some ways of
escaping from this somewhat disconcerting conclusion.

\section{Einstein conformal scalar field solutions}
\label{e}

In this section, we will describe the exact solutions to the combined
system of equations for Einstein gravity and a massless conformally
coupled scalar field, in the simple case of static and spherically
symmetric configurations.

The Einstein equations can be written as $\kappa \;
G_{\mu\nu}=T_{\mu\nu}$, where $G_{\mu\nu} = R_{\mu\nu} - \frac{1}{2}
g_{\mu\nu}\;R$ is the Einstein tensor, with $R_{\mu\nu}$ the Ricci
tensor and $R$ the scalar curvature. $T_{\mu\nu}$ is the
energy-momentum tensor of the matter field, and $\kappa=( 8\pi
G_N)^{-1}$, with $G_N$ denoting Newton's constant.  For a massless
conformal scalar field $\phi_c$, the energy-momentum tensor acquires
the form \cite{flanagan}
\begin{equation}
T_{\mu\nu} = \nabla_{\mu} \phi_c \nabla_{\nu} \phi_c 
- {1 \over 2} g_{\mu\nu} (\nabla \phi_c)^2 
+ {1 \over 6} \left[ G_{\mu\nu} \; \phi_c^2 
- 2 \;\nabla_{\mu} ( \phi_c \nabla_{\nu} \phi_c) 
+ 2 \; g_{\mu\nu} \; \nabla^{\lambda} (\phi_c \nabla_{\lambda} \phi_c) 
\right],
\end{equation}
with the field satisfying the equation $(\Box - {1\over6}R) \phi_c=0$.
This is the generalization to curved spacetime of the ``new improved
energy-momentum tensor'' more usually invoked in a particle physics
context \cite{callan}.

The key feature of the energy-momentum tensor for a conformal field is
that it is traceless, $T_{\mu\nu} \; g^{\mu\nu}=0$, and therefore
$R=0$.  For this reason, we can write the coupled Einstein plus {\em
conformal} scalar field equations as
\begin{eqnarray}  
R_{\mu\nu}=&&\hspace*{-6mm}\left(\kappa-{1 \over 6}\phi_c^2 \right)^{-1}
\left( 
{2 \over 3} \nabla_{\mu} \phi_c \nabla_{\nu} \phi_c 
-{1 \over 6} g_{\mu\nu} (\nabla \phi_c)^2 
-{1 \over 3}\phi_c \nabla_{\mu}\nabla_{\nu} \phi_c
\right),  
\label{00} 
\\ 
\Box\phi_c=&&\hspace*{-6mm}0. \label{laplacian}
\end{eqnarray}
We are interested in static and spherically symmetric solutions of
these equations. In order to find these solutions, we will start by
looking for metrics conformally related to the
Janis--Newman--Winicour--Wyman (JNWW) \cite{jnw,wyman,virbhadra}
static spherically symmetric solutions of the Einstein {\em minimally
coupled} massless scalar field equations:
\begin{eqnarray}  
R_{\mu\nu}=&&\hspace*{-6mm}\kappa^{-1}
\left( 
\nabla_{\mu} \phi_m \nabla_{\nu} \phi_m 
\right),  \label{00m} 
\\ 
\Box\phi_m=&&\hspace*{-6mm}0. \label{laplacianm}
\end{eqnarray}
The JNWW solutions can be expressed \cite{agnese} as 
\begin{eqnarray}  
&&\hspace*{-8mm}ds_m^2=-\left(1-{2\eta \over r} \right)^{\cos\chi} dt^2
+\left(1-{2\eta \over r} \right)^{-\cos\chi} dr^2
+\left(1-{2\eta \over r} \right)^{1-\cos\chi} 
r^2 (d\theta^2+\sin^2 \theta \; d\Phi^2), 
\nonumber  \\ 
&&          \\
&&\hspace*{+3.6cm}\phi_m=\sqrt{{\kappa \over 2 } }\; \sin \chi \;  
\ln \left(1-{2\eta \over r} \right).
\end{eqnarray}
The JNWW solutions possess obvious symmetries under $\chi\to -\chi$,
(with $\phi_m \to -\phi_m$). Less obvious is that by making a
coordinate transformation $r\to\tilde r = r - 2\eta$, one uncovers an
additional symmetry under $\{ \eta,\chi \} \to \{ -\eta,\chi+\pi
\}$, (with $\phi_m \to +\phi_m$). In view of these symmetries one can
without loss of generality take $\eta \geq 0$ and $\chi \in
[0,\pi]$. Similar symmetries will be encountered for conformally
coupled scalars.\footnote{
The key to this symmetry is to realise that
\[
\left( 1 - {2\eta\over r} \right) = 
\left( 1 + {2\eta\over \tilde r} \right)^{-1}.
\]}

The requirement that a metric conformal to the JNWW metric,
$ds=\Omega(r) \; ds_m$ with $\Omega(r)$ the conformal factor, have a
zero scalar curvature [necessary if it has to be solution of the
system of equations (\ref{00})], easily provides a second-order
differential equation for the conformal factor as a function of
$\phi_m$:
\begin{equation}
{d^2 \Omega(\phi_m) \over d {\phi_m}^{2} } = 
{1 \over 6 \kappa} \; \phi_m.
\end{equation}
Its solutions can be parametrized in the form
\begin{equation}
\Omega=\alpha_{+}\;{\rm exp}\left(+\phi_m / \sqrt{6 \kappa } \right) 
+\alpha_{-}\;{\rm exp}\left(-\phi_m / \sqrt{6 \kappa} \right),
\end{equation}
with $\alpha_{+}$ and $\alpha_{-}$ two real constants.
The equation $\Box\phi_c=0$ can now be integrated yielding
\begin{equation}
\phi_c = A \;{\sqrt{6 \kappa } \over 4 \alpha_{+}\alpha_{-}}
\left[{ \alpha_{+}{\rm exp}\left(+\phi_m / \sqrt{6 \kappa } \right) 
-\alpha_{-}{\rm exp}\left(-\phi_m / \sqrt{6 \kappa} \right)
\over   \alpha_{+}{\rm exp}\left(+\phi_m / \sqrt{6 \kappa } \right) 
+\alpha_{-}{\rm exp}\left(-\phi_m / \sqrt{6 \kappa} \right) 
}\right] +B,
\end{equation}
where $A$ and $B$ are two additional integration constants.

In order that the metric and conformal scalar field just found be
solutions of the whole set of equations (\ref{00}), the four
integration constants, $\alpha_{+}$, $\alpha_{-}$, $A$, and $B$, must
be inter-related in a specific way.  After a little algebra the
equation for the $tt$ component in (\ref{00}) implies the
following relations\footnote{
In view of the assumed static and spherical symmetries, the Einstein
equations provide only three constraints, and by the contracted
Bianchi identities only two of these are independent. Thus it is only
necessary to consider the trace $R$ and the $tt$ component $R_{\hat t
\hat t}$ to guarantee a solution of the entire tensor equation.}
%
\begin{eqnarray}  
&&\hspace*{6mm}A \; B \; \alpha_{+}\; \alpha_{-}=0, \\
&&\hspace*{-6mm}\alpha_{+}^2 \; \alpha_{-}^2 \; 
\left(1-{B^2 \over 6 \kappa} \right)
+{A^2 \over 16}=0.
\end{eqnarray}
Therefore, we have two options: 

{\em Case i}) $B=0$, $A=\pm 4 \; \alpha_{+} \; \alpha_{-}$; \\
\begin{eqnarray}
&&\hspace*{-6mm}\Omega=\alpha_{+}\;{\rm exp}
\left(+\phi_m / \sqrt{6 \kappa } \right) 
+\alpha_{-}\;{\rm exp}\left(-\phi_m / \sqrt{6 \kappa} \right), \\
&&\hspace*{-6mm}\phi_c=\pm\sqrt{6 \kappa }
\left[{ \alpha_{+}\;{\rm exp}\left(+\phi_m / \sqrt{6 \kappa } \right) 
-\alpha_{-}\;{\rm exp}\left(-\phi_m / \sqrt{6 \kappa} \right)
\over   \alpha_{+}\;{\rm exp}\left(+\phi_m / \sqrt{6 \kappa } \right) 
+\alpha_{-}\;{\rm exp}\left(-\phi_m / \sqrt{6 \kappa} \right)
}\right].
\end{eqnarray}

{\em Case ii}) $A=0$, $B=\pm \sqrt{6 \kappa }$;
\begin{eqnarray}
&&\hspace*{-6mm}\Omega=\alpha_{+}\;{\rm exp} 
\left(+\phi_m / \sqrt{6 \kappa }\right) 
+\alpha_{-}\;{\rm exp}\left(-\phi_m / \sqrt{6 \kappa} \right),
\\
&&\hspace*{-6mm}\phi_c=\pm\sqrt{6 \kappa }.
\end{eqnarray}
Notice that these two branches of solutions intersect when either
$\alpha_{+}$ or $\alpha_{-}$ are equal to zero. 

The set of solutions we have found is a generalization of the
solutions found by Froyland \cite{froyland}, and some time later by
Agnese and La Camera \cite{agnese}. Indeed, in the case
$\alpha_{+}=\alpha_{-}$ the conformal factor becomes $\Omega = \cosh(
\phi_m / \sqrt{6 \kappa })$, (we drop a unimportant constant factor),
and the field $\phi_c =
\pm \sqrt{6 \kappa } \; \tanh( \phi_m / \sqrt{6 \kappa })$, in agreement
with the expression given by Agnese and La Camera.  The Froyland
solution is in fact identical to that of Agnese--La Camera though this
is not obvious because Froyland chose to work in Schwarzschild
coordinates. Because of this, Froyland could only provide an implicit
(rather than explicit) solution.\footnote{
That is, Froyland calculated ${\cal R}(\phi_c)$, which in Schwarzschild
coordinates is not analytically invertible to provide $\phi_c({\cal
R})$. The coordinate system chosen by Agnese and La Camera is much
better behaved in this regard and $\phi_c(r)$ can be explicitly
calculated as we have seen above. The trade-off is that whereas ${\cal
R}(r)$ can be written down explicitly, there is no way of analytically
inverting this function to get $r({\cal R})$.}

Let us now analyze the different behaviours of these solutions.  For
this task, it is convenient to look at the conformal factor as a
function of $r$,
\begin{equation}
\Omega(r)=
\alpha_{+}\left(1-{2\eta \over r} \right)^{{\sin\chi \over 2 \sqrt{3}}}
+\alpha_{-}\left(1-{2\eta \over r} \right)^{-{\sin\chi \over 2 \sqrt{3}}}.
\end{equation}
In the same way, as a function of $r$, the Schwarzschild radial
coordinate ${\cal R}$ has the form
\begin{equation}
{\cal R}(r)=r\left(1-{2\eta \over r} \right)^{{1-\cos\chi \over 2}}\Omega(r).
\end{equation}  

If we define an angle $\Delta$ by $\tan(\Delta/2)=
(\alpha_{+}-\alpha_{-}) / (\alpha_{+}+\alpha_{-})$, then the domain
$\Delta \in (-\pi, \pi]$ exhausts all possible metric configurations,
as a constant overall factor in the metric can be absorbed in the
definition of coordinates. We have a three-parameter family of
solutions depending on $\eta$, the angle $\Delta$, and the angle $\chi
\in (-\pi, \pi]$. In fig. 1 we have drawn the parameter space as a 
square. Indeed, parallel edges are identified, so the parameter space
is an orbifold (a two-torus subjected to symmetry identifications).
In fact, the solution space is invariant if we change $\{ \eta, \chi,
\Delta \}$ to $\{ \eta, -\chi, -\Delta \}$. Furthermore, the coordinate
transformation $r \to \tilde r = r - 2\eta$ can now be used to deduce
an invariance under $\{ \eta, \chi, \Delta \} \to \{ -\eta, \chi+\pi,
\Delta \}$. Combining the two symmetries, we deduce an invariance
under $\{ \eta, \chi, \Delta \} \to \{ -\eta, \pi-\chi, -\Delta
\}$. Thus without loss of generality we only have to deal with the
region $\eta\geq 0$ with $\chi \in [0, \pi]$. Also, we need only
consider the geometries in which $\Delta \neq \pi$ because, for that
value, the geometries do not have an asymptotic flat region when $r$
approaches infinity.\footnote{
We mention in particular that the physical scale length, effective
Newton constant, and physical mass of the spacetime can be isolated
from a weak-field expansion near spatial infinity. We find
\[
G_{\mathrm{eff}} \; M = \eta \; \left( \cos\chi + 
\tan\left[{\Delta\over2}\right] \; {\sin\chi\over\sqrt{3}} \right).
\]
We note that this scale length is invariant under all the symmetries
mentioned above (as it must be). The effective Newton constant is
\[
G_{\mathrm{eff}} 
= {1\over 8\pi ( \kappa - {1\over6} \phi_\infty^2 ) } 
= {1\over8\pi \kappa} \left(1-\tan^2{\Delta\over2}\right)^{-1}.
\]
Thus $G_{\mathrm{eff}}$, $M$, and $G_{\mathrm{eff}}\; M$ are separately
invariants of the symmetries discussed above. For the objectional
value $\Delta=\pi$, the lack of an asymptotically flat region is
reflected in an infinite physical mass.}

\begin{figure}[htb]
\vbox{ 
\hfil
\scalebox{\SIZE}{\includegraphics{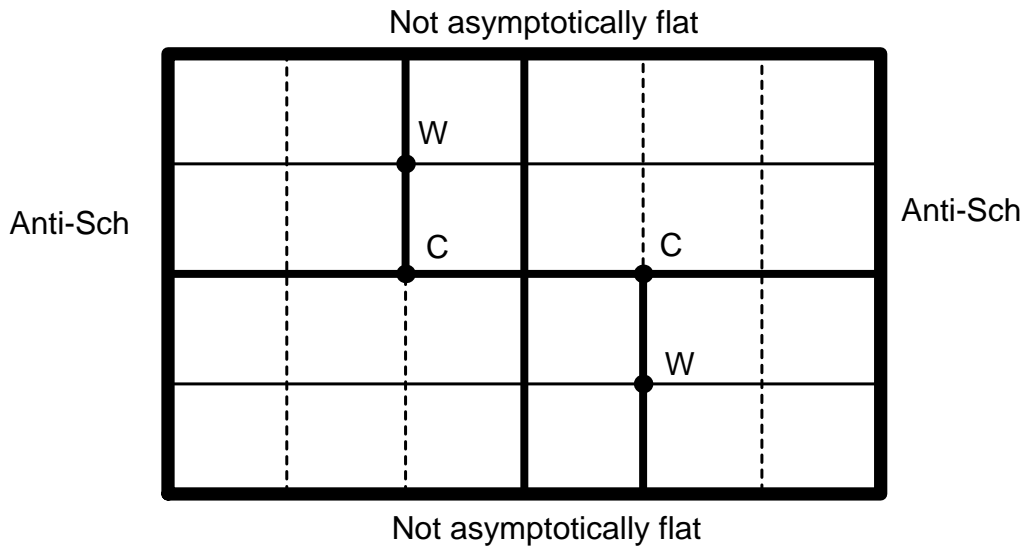}}
\hfil 
}
\bigskip
\caption{\label{F:phase} 
Sketch of the parameter space of our class of solutions to Einstein
gravity coupled to the ``new improved stress-energy tensor''. Using
the symmetries discussed in the text we have taken $\eta\geq 0$. We
sketch the range $\chi \in [-\pi,+\pi]$ and $\Delta \in [-\pi,+\pi]$.
Parameter space is an orbifold because of the remaining symmetry under
$\{ \eta, \chi,\Delta \}$ to $\{ \eta, -\chi, -\Delta \}$. The left
and right borders of the figure correspond to anti-Schwarzschild space
(the negative mass Schwarzschild geometry), while the top and bottom
borders correspond to spacetimes that are not asymptotically flat. The
central vertical line corresponds to the Schwarzschild black hole,
while the central horizontal line represents the Froyland--Agnese--La
Camera branch of solutions. The two offset vertical branches are the
special class of traversable wormholes with two asymptotically flat
regions that are the main focus of this paper. These branches
terminate in ``cornucopia'' labelled $C$, and contain the special
symmetric traversable wormholes labelled $W$.}
\end{figure}

For the rest of parameter space, an analysis when $r$ approaches
$2\eta$ of the behaviour of the $tt$ component of the Ricci tensor
($R_{\hat t\hat t}$ in an orthonormal coordinate basis) shows that it
diverges for every point in parameter space except when $\chi=0$, $\{
\chi={\pi \over 3}, \Delta \neq {\pi \over 2} \}$, $\{ \chi={2\pi
\over 3}, \Delta={\pi \over 2} \}$, or $\chi=\pi$.  Thus, except for
these parameter values, we have geometries with a naked curvature
singularity at $r=2\eta$.

Among these singular geometries, those with $0<\chi<{\pi \over 3}$ and
$\Delta \neq {\pi \over 2}$ deserve additional attention. For them,
the Schwarzschild radial coordinate blows up when approaching
$r=2\eta$.  They are geometries with wormhole shape, but with
something ``strange'' in the other ``asymptotic'' region. We can
easily check that the proper radial distance between every point
$r>2\eta$ and $r=2\eta$ is finite. Also the proper volume beyond every
sphere at finite $r>2\eta$ is itself finite, even though the proper
area of the spherical sections diverges as one approaches $r=2\eta$.
Therefore, although one certainly encounters a flare-out (a wormhole
throat) before reaching $r=2\eta$, we cannot speak properly of
another asymptotic region.\footnote{
Thus these geometries are certainly Lorentzian wormholes, and could
even be called ``traversable in principle'', but because of the nasty
behaviour on the other side of the wormhole throat, they do not deserve
to be called ``traversable in practice''.}

The first non singular case is $\chi=0$.  In this case we recover the
Schwarzschild black hole geometry.  This geometry, of course, does not
have a curvature singularity at $r=2\eta$, just a coordinate
singularity.  It is instead singular at $r=0$.

For the special cases $\{ \chi={\pi \over 3}, \Delta \neq {\pi \over
2} \}$ there exist situations in which the Schwarzschild radial
coordinate ${\cal R}$ and the gravitational potential $g_{tt}$ go to
non-zero constants when approaching $r=2\eta$. This suggests that we
might be able to extend the geometry beyond $r=2\eta$. We leave for
the next section the analysis of how this extension can be done,
showing that there exist genuine wormhole solutions.

In the case $\{ \chi={2\pi \over 3}, \Delta={\pi \over 2} \}$ we can 
write the metric as 
\begin{equation}
ds^2=-dt^2+\left(1- {2\eta \over r} \right) dr^2+ (r-2\eta)^2 
(d\theta^2+\sin^2\theta \; d\Phi^2).
\end{equation} 
Writing $\tilde r=r-2\eta$, so that
\begin{equation}
ds^2=-dt^2+ {d{\tilde r}^2 \over \left(1+ {2\eta \over \tilde r} \right) }
+ {\tilde r}^2 (d\theta^2+\sin^2\theta \; d\Phi^2),
\end{equation} 
a brief calculation shows that this geometry is also singular at
$\tilde r=0$, $r=2\eta$, even though $R_{\hat t\hat t}$ remains finite
there. (The other Ricci components, $R_{\hat r\hat r}$ and
$R_{\hat\theta \hat\theta}$, diverge as $\tilde r\to 0$.)

Finally, the case $\chi=\pi$ corresponds to a negative mass
Schwarzschild geometry. It has a naked curvature singularity at
$r=2\eta$, but after the coordinate change $r\to\tilde r = r-2\eta$,
the naked curvature singularity moves to $\tilde r=0$, and the
character of the geometry becomes obvious.

\section{Traversable wormhole solutions}
\label{wh}

This section is devoted to the analysis of the case 
$\{ \chi={\pi \over 3}, \Delta \neq {\pi \over 2} \}$.
For this task, it is convenient to change to isotropic coordinates
\begin{equation}
r=\bar r \left(1+{\eta \over 2  \bar r}   \right)^2.
\end{equation}   
With the new radial coordinate running from $\bar r=\eta/2$ to
$\infty$ we cover the same portion of the metric manifold that $r \in
[\eta, \infty)$ did before. However, can the manifold be analytically
extended beyond $\bar r=\eta/2$? The answer is yes.  We can write the
metric in isotropic coordinates as
\begin{equation}
ds^2=
\left[
\alpha_{+}
{\left(1-{\eta \over 2\bar r} \right) 
\over 
\left(1+{\eta \over 2\bar r} \right)}    
+ 
\alpha_{-}   
\right]^2
\left[ - dt^2 + \left( 1+{\eta \over 2  \bar r}   \right)^4 [d{\bar r}^2+
{\bar r}^2 (d\theta^2+\sin^2\theta \; d\Phi^2)]      \right],
\label{wormholes}
\end{equation}   
noticing that it is perfectly well behaved at $\bar r = \eta/2$.  We
want to point out that for $0 < \bar r < \eta/2$ the conformal factor
$\Omega^2$ is real and negative, while the JNWW solution is
ill-behaved in the sense that the metric $ds^2_m$ has opposite
signature to the usual. Nevertheless, the metric $ds^2$ is perfectly
well behaved. Thus, strictly speaking, only the region $\bar r >
\eta/2$ is conformally related with its corresponding JNWW solution.

Among these solutions, in those with $\Delta \in (-\pi, 0)$ the
Schwarzschild radial coordinate ${\cal R}(\bar r)$ reaches a minimum
value at $\bar r= (\eta/2) \; |\tan(\Delta/2)|^{1/2}$, and comes back
to infinity at $\bar r=0$, showing up as another asymptotically flat
region. As well as this, the $tt$ component of the metric is
everywhere non-zero. Therefore, the region in parameter space $\{
\chi= {\pi \over 3}, \Delta \in (-\pi, 0)\}$ represents genuine
traversable wormhole solutions, with the throat of the wormhole being
located at $\bar r= (\eta/2) \; |\tan(\Delta/2)|^{1/2}$.  For
completeness, in the solution with $\Delta=0$, ${\cal R}(\bar r)$
reaches a minimum at $\bar r=0$, but this sphere is at a infinite
proper distance from every other $\bar r$ so we can conclude that this
geometry is a ``cornucopia'' (tube without end). ($\Delta=-\pi$
represents the reversed cornucopia with no asymptotic region at $\bar
r=\infty$).  For the remaining values of $\Delta \in (0,\pi)$ the
geometry pinches off (the conformal prefactor in equation
(\ref{wormholes}) goes to zero) at finite positive radius $\bar r =
(\eta/2) \; \tan(\Delta/2)$.

We have seen that a scalar field coupled conformally to gravity can 
support wormhole geometries. On these wormhole configurations the 
conformal scalar field takes the following form
\begin{equation}
\phi_c=\pm\sqrt{6 \kappa} \; 
{\alpha_{+} \left( 1-{\eta \over 2 \bar r} \right)
-\alpha_{-} \left( 1+{\eta \over 2 \bar r} \right) \over 
\alpha_{+} \left( 1-{\eta \over 2 \bar r} \right)
+\alpha_{-} \left( 1+{\eta \over 2 \bar r} \right)  }.
\end{equation}  
It is a monotonically increasing or decreasing function (depending 
on the overall sign) between one asymptotic region and the other, 
taking the values
\begin{equation}
\phi_c|_{\mathrm{asym_1}}
=\pm\sqrt{6 \kappa} \;\tan{\Delta \over 2}, 
\hspace{0.5cm} {\rm and} \hspace{0.5cm} 
\phi_c|_{\mathrm{asym_2}}
=\pm {\sqrt{6 \kappa} \over \tan{\Delta \over 2}}.
\end{equation} 
This monotonic behaviour for the scalar field causes an asymmetry
between the asymptotic regions. In fact, we can realize of the real
importance of this asymmetry by looking at the effective Newton's
constant, $G_{{\rm eff}}=8 \pi (\kappa- {1 \over 6} \phi_c^2)^{-1}$,
that can be defined on these systems.  It not only reaches a different
value on each asymptotic region,
\begin{equation}
G_{{\rm eff}}|_{\mathrm{asym_1}}
={1 \over 8 \pi \kappa} {1 \over (1-\tan^2 {\Delta \over 2}) },
\hspace{1cm} 
G_{{\rm eff}}|_{\mathrm{asym_2}}
=-{1 \over 8 \pi \kappa} {\tan^2 {\Delta \over 2} \
\over (1-\tan^2 {\Delta \over 2}) },
\end{equation} 
it also reaches a different {\em sign}. From the point of view of the
asymptotic region with a positive effective Newton's constant, the
wormhole throat is located in the region in which $G_{{\rm eff}}$ has
already changed its sign.  This asymmetry is reflected also in the
values of the asymptotic masses measured on the two sides of the
wormhole throat. The scale lengths are
\begin{equation}
(G_{{\rm eff}} M)|_{\mathrm{asym_1}}
={\eta \over 2} 
\left(1+\tan{\Delta \over 2} \right),
\hspace{1cm} 
(G_{{\rm eff}} M)|_{\mathrm{asym_2}}
={\eta \over 2} 
\left(1+{1 \over \tan{\Delta \over 2} } \right).
\end{equation}
The observers in the asymptotic region with a positive $G_{{\rm eff}}$
see a positive asymptotic mass, while those in the other asymptotic
region see a negative $G_{{\rm eff}}$ and a positive asymptotic mass.
The case $\Delta=-{\pi \over 2}$ corresponds to $\alpha_{+}=0$, so
\begin{eqnarray}
&&\hspace{-6mm}ds^2=-dt^2+\left(1+ {\eta \over 2 \bar r} \right)^4 
[d{\bar r}^2+ {\bar r}^2(d\theta^2+\sin^2\theta \; d\Phi^2)], 
\label{symwh} \\
&&\hspace{3cm}\phi_c=\pm \sqrt{6 \kappa}.
\end{eqnarray}
This describes a limiting symmetric wormhole with an everywhere zero
effective Newton's constant and both asymptotic masses equal to zero.

These asymmetric features are disappointing because we would like to
have wormholes connecting equivalent regions of space. At this point
we have to say that while a conformal scalar field can provide us with
the ``flaring out'' condition for geodesics, it has the drawback of
reversing the sign of the effective Newton's constant in the other
asymptotic region.

\section{Discussion}
\label{d}

We have found that, among the classical solutions of general
relativity coupled to a massless conformal scalar field there exist
genuine Lorentzian traversable wormhole geometries. Although perfectly
well behaved from a geometrical point of view, they are asymmetric in
the sense that the effective Newton's constant has a different sign in
each asymptotic region.

Inspecting the expression for the Laplacian of a scalar field in a
static and spherically symmetric geometry
\begin{equation}
\Box \phi_c={1 \over \sqrt{-g} } \; \partial_r
\left[\sqrt{-g} \; g^{rr} \; \partial_r \phi_c \right],
\end{equation}
we see that in order for the scalar field to be able to change its
monotonic behaviour in a non-singular geometry, there must be some
points at which $\Box \phi_c\neq 0$.  This suggests that in more
general situations that those analyzed in this paper we could find
traversable wormhole solutions with no asymmetry between the
asymptotic regions.  For example, if we add to our system a quantity
of normal matter with a positive trace for the energy-momentum tensor
($T>0$), and we place this normal matter in two thin spherical shells
(see fig. 2), we can join smoothly an inner region with the geometry
of the symmetric wormhole solution (\ref{symwh}) to two outer
asymptotic geometries, both with positive effective Newton's
constants.\footnote{
Thus we are considering geometries that are built piecewise out of
segments of the solutions considered in this paper, with junction
conditions applied at the thin shells.}
The requirement that $T>0$ in the shells translates into a localized
negative scalar curvature, $R<0$, necessary for bringing down the
value of the scalar field from that in the inner region.

Finally, we conclude by emphasising that it is not so much the
occurrence of classical wormholes in and of themselves that is the main
surprise of this paper.  Rather, what is truly surprising here is that
such an inoffensive and physically well-motivated classical source,
the ``new improved stress-energy tensor'', leads to classical
traversable wormholes. 

\begin{figure}[htb]
\vbox{ 
\hfil
\scalebox{1.00}{\includegraphics{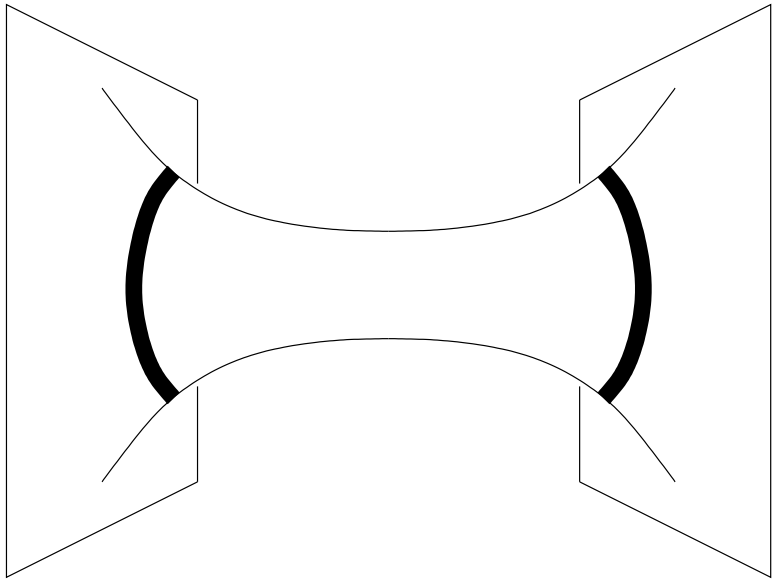}}
\hfil 
}
\bigskip
\caption{\label{F:shell} 
Schematic depiction of how two thin shells of ordinary matter could be
combined with the flare out effect coming from the ``new improved
stress-energy tensor'' to build a traversable wormhole that has nice
asymptotic properties on both sides of the wormhole throat. The
geometry is piecewise a solution of the field equations discussed in
this paper, with a thin shell of normal stress-energy being used to
start the scalar field moving downwards from $\phi_c =
\sqrt{6\kappa}$.}
\end{figure}

\begin{figure}[htb]
\vbox{ 
\hfil
\scalebox{1.00}{\includegraphics{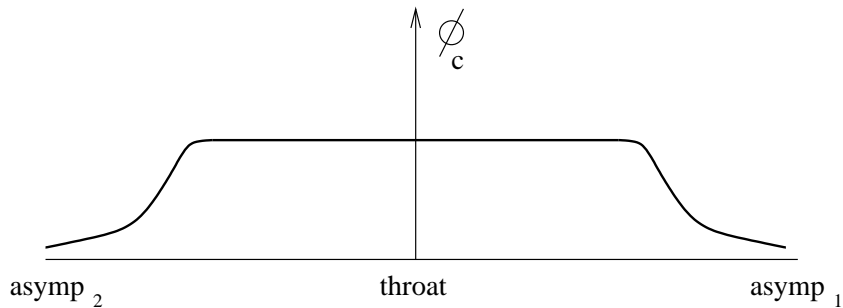}}
\hfil 
}
\bigskip
\caption{\label{F:field} 
Schematic depiction of the scalar field $\phi_c$ as a function of
position as one traverses the throat of this ``cut and paste''
wormhole.}
\end{figure}

\section*{Acknowledgments}

The research of CB was supported by the Spanish Ministry of Education
and Culture (MEC). MV was supported by the US Department of Energy.

 
\end{document}